# Data-Mining Element Charges in Inorganic Materials

Yu Ding[a], Yu Kumagai[b], Fumiyasu Oba[b] and Lee A. Burton[a]*

[a] International Centre for Quantum and Molecular Structures, Department of Physics, Shanghai University, Shanghai 200444, China.

[b] Laboratory for Materials and Structures, Institute of Innovative Research, Tokyo Institute of Technology, 4259 Nagatsuta, Midori-ku, Yokohama 226-8503, Japan.

AUTHOR INFORMATION

**Corresponding Author**

* E-mail: leeburton@shu.edu.cn Fax: +86 21 66134208 ; Tel: +86 2166136129;

ABSTRACT Oxidation states are well-established in chemical science teaching and research. We data-mine more than 168,000 crystallographic reports to find an optimal allocation of oxidation states to each element. In doing so we uncover discrepancies between text-book chemistry and reported charge states observed in materials. We go on to show how the oxidation states we recommend can significantly facilitate materials discovery and heuristic design of novel inorganic compounds.

# TOC GRAPHICS

Insert 2in×2in (5cm×5cm) graphic

The graphic for the TOC/Abstract should representative of your entire work and not be a duplicate of a graphic already used in the manuscript. Color structures, graphical images, photographs, or reaction schemes are typically good choices.

Computer models have become powerful enough to predict previously unknown compounds entirely from first principles.[1] While such a feat is impossible from a purely statistical standpoint,[2,3] it has proven feasible with the judicious use of chemical heuristics to restrain the combinatorial explosion of possibilities. One such axiom is the principle of electroneutrality, *i.e* while species within a material may have a local charge accumulation or depletion, overall the material's composition must correspond to a net total charge of zero.

The extent to which assigned charges, so-called oxidation states, are meaningful has been debated for more than 200 years,[4] because the charge of a given ion is not directly observable,[5,6] and exceptions to the rule exist. [7] Furthermore, oxidation states are quite flexible, any pure element has a formal oxidation of zero but the highest known oxidation state is reported to be +9 for Ir in the $IrO_4^+$ complex,[8] and even +10 for Pt in the cation $PtO_4^{2+}$.[9] Even so, the conceptual assignment of oxidation states within a material is prevalent enough that their use is extolled for introductory level teaching even by the International Union of Pure and Applied Chemists (IUPAC).[10]

We scan an inorganic crystal structure database (ICSD) repository of 169,800 entries, which overwhelmingly correspond to materials that have been successfully synthesized and characterized in experiment. In this letter, we show that despite a wide range of *possible* oxidation states, the range of *probable* oxidation states is in fact rather narrow. We search for 108 elements between -4 and +8 assigned charge, making 1,296 possible element-charge states in total and find the vast majority of reports correspond to only 84 oxidation states in total.

The recovered oxidation states relative to the total number of reports for that element are plotted in Figure 1a. The apparent trends observed in oxidation state as a function of element number reflect the valence electrons' orbital quantum number, with clear trends for *s*,*p*,*d* and *f* shell behavior, just like the equivalent 'blocks' of the periodic table. Abrupt transitions from largely positive to negative charge states correspond to change in tendency from depleting a partially full electron shell, to completing the same shell. The ambivalent nature of the early d block, which is widely utilized for redox chemistry and catalysis is observable, as is the fairly uniform ionic behavior of the heavier f-elements. Thus, oxidation states can provide an instructive cross-section of element behavior in inorganic chemistry.

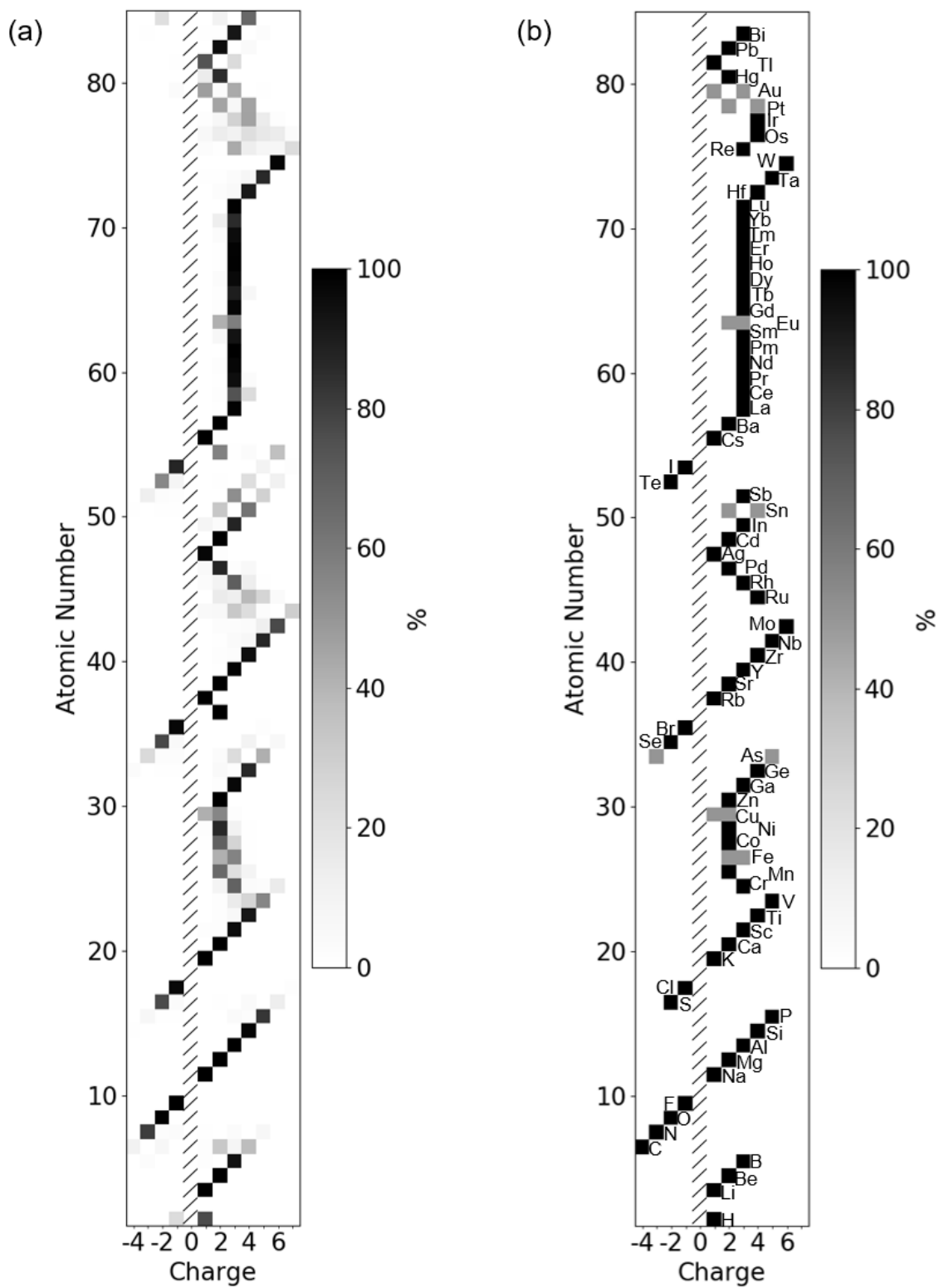

Figure 1: (left) the oxidation states for each element reported in the ICSD as a percent of all reports for that element (right) the 84 proposed oxidation states that capture more than 90 % of all reports in the ICSD. Only the right figure is labelled to prevent obstruction of the finer detail in the ICSD data on the left. Charge zero has no data in either plot and is represented by a strikethrough. The raw data for both plots is provided in the SI.

The normalization to each element obscures the fact that some are overwhelmingly more common than others. Factors such as abundance, cost, toxicity, radioactivity *etc.* all affect the willingness of a scientist to work with an element and these are naturally reflected in the popularity of reports. Thus, while both Sr and Kr are reported as 100 % 2+, there are more than 400 times the number of Sr reports than Kr. In considering the total of oxidation states overall, we find that the vast majority of reported chemistries can be recovered with 84 representative oxidation states, which we show in Figure 1b. All reports greater than 40 % of the relative reports for that element up to Polonium are covered by the 84 oxidation states we propose, excluding Tc, Kr and Xe that we deliberately exclude. The most frequent reported oxidation state we don't include is hydride ion ($H^{-1}$), with more than 20,000 reports, opting instead for $H^{+1}$ with almost 68,00 reports. The largest oxidation state relative to that element we don't include is $C^{4+}$. We propose the state 4- as the representative oxidation state of carbon because it is the most common among the negative oxidation states and carbon in inorganic chemistry is often sought in the context of carbide materials, e.g. tungsten (IV) carbide, which has many useful properties for industry.[11] We attribute that the popularity of the $C^{4+}$ assignment to organic molecules that are present even in the inorganic crystal structure database in metal organic frameworks and hybrid materials, see for example methyl ammonium lead iodide.[12] Such distinct behavior helps validate the pedagogical approaches used in chemistry, in the division between organic and inorganic behavior, even in solids state. Overall however, more than 90 % of all assigned oxidation states in our repository that are not zero (1,534,032 of 1,698,849) are captured by our 84 states, see Figure 2.

There are other recommendations for oxidation states in the literature. The CRC Handbook has 184 oxidation states for 98 elements (excluding elements that have oxidation 0 or none).[13] However, we find some of these states are hardly reported in the ICSD. For example, there are only 70 instances of 4+ nitrogen out of a total of 69,590 charge assignments for that element. It is perhaps the case that such anomalous states are included as possible rather than probably oxidation states. However, the textbook The Chemistry of Elements distinguishes between possible and probable oxidation states, and, if anything, this gives rise to even more issues.[14] This textbook has 367 oxidation states for 101 elements (excluding elements that have oxidation 0 or none), 157 of which are denoted as common oxidation states. Even the oxidation states denoted as common are overstated compared to the number we recover statistically. For example, $S^{2+}$, $Cl^{1+}$ and $Si^{4-}$ are all listed as a common oxidation states but in fact, account for less than 1 % of the instances of oxidation state assignments for those elements in the ICSD. Furthermore, there are several cases of possible oxidation states listed that have zero reports in the ICSD, including but not limited to $Li^{1-}$, $O^{2+}$, $Cl^{6+}$, $Ti^{1-}$, $V^{1-}$, $Mo^{2-}$, $Nb^{1-}$, $Ru^{2-}$, $Rh^{6+}$, $Ta^{1-}$, $W^{1-}$, $Re^{3-}$. Finally, there are mis-identified common oxidation states, where a charge is reported to be more common than others but in fact appears less often than alternatives in the ICSD. Some of these are: $Mn^{4+}$ as more common than $Mn^{3+}$, $Re^{4+}$ as more common than $Re^{3+}$ and $Au^{3+}$ as more common than $Au^{1+}$ in the textbook, where the inverse is true in the ICSD.

We believe our 84 states not only better reflect the chemistry of solid systems but by being smaller in number significantly streamline viable possibilities for structure prediction and materials design. The majority of structure prediction methods rely, or at least are accelerated by a provided composition, be it with ionic substitution,[15] genetic algorithm,[16] random structure search,[17] or machine learning approaches.[18]

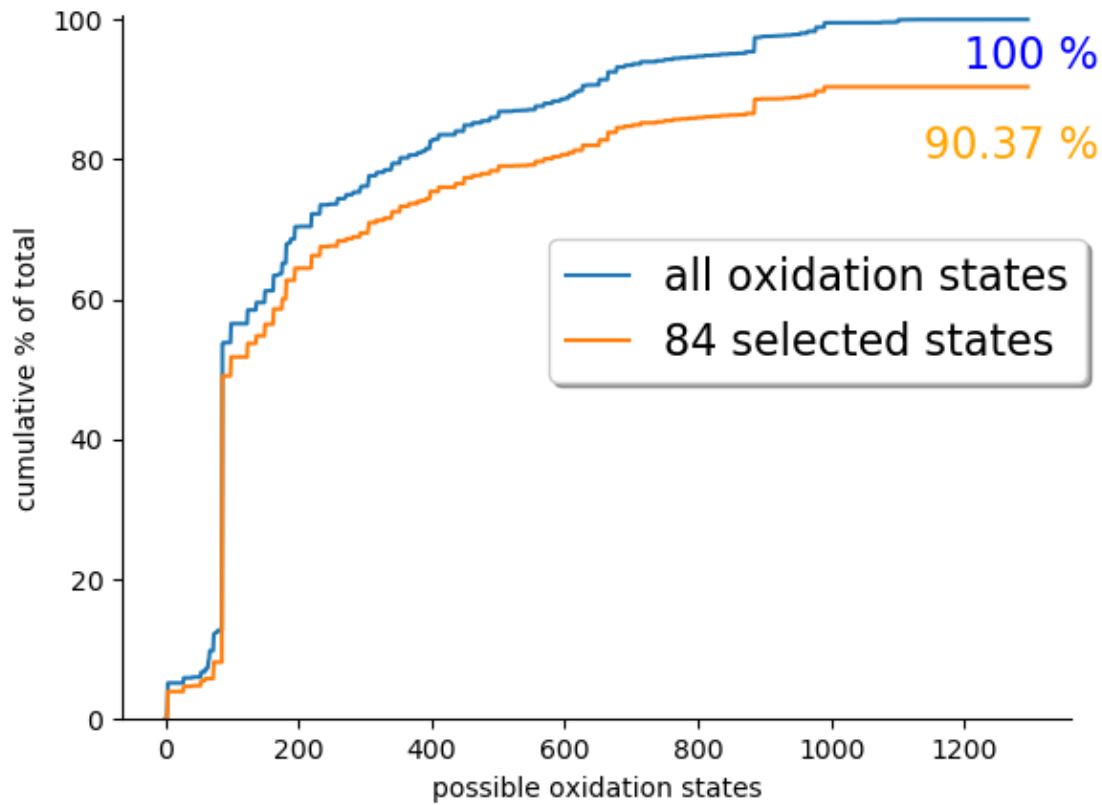

Figure 2: (blue) cumulative percent of total matches across all 1,296 possible oxidation states (-4 to +8 for each element excluding zero). (Orange) cumulative percent of oxidation states recovered by a selection of 84 states across the same data.

To illustrate the degree to which our results can facilitate materials discovery, we prepare a simple algorithm that combines oxidation states together and returns only those that sum to zero *i.e.* comply with the principle of electroneutrality. We provide an example implementation of such a script in the SI as text and online as an executable.[1] It approximates the number of viable chemical compositions from component elements *via* combination with repetition *i.e.* $\binom{n+k-1}{k}$

---

[1] https://github.com/bmd-lab/electroneutral_match

for a selection of $k$ components from a set of $n$ oxidation states. The repetition is necessary to enable multiple of the same component within a given composition, or formula, as with Cu in $Cu_2S$ or S in $SnS_2$. By using combinations, instead of permutations, the order of the elements is ignored, such that SnS is not double counted as SSn, for example.

The results of using this method with our list of 84 oxidation states, the textbook common and textbook possible oxidation states, are shown in Table 1 (we also provide these 3 sets of oxidation states as files online). It can clearly be seen that including additional oxidation states comes at an extreme penalty in terms of candidates to be considered, even with enforced electroneutrality.

While we only report up to 5 components in Table 1, to identify a compound such as the well-studied $Cu_2ZnSnS_4$ (CZTS) *a priori*, without employing any chemical heuristics, would require a selection of eight components (k =8), from the full set of oxidation states. Even an algorithm as rudimentary as the one we provide becomes impossible to run at such a high number of components for the larger sets of oxidation states. It may seem extreme to include CZTS but it is one of many materials with such a high number of components that are of interest to the community, see for example double perovskites, that are receiving a significant amount of attention currently and have chemical formulae with 10 or more components.[19,20] For this reason, the principle of electroneutrality has been considered insufficient to reduce the combinatorial explosion sufficiently to render *a priori* materials discovery viable, and other approximations have also been employed. For example, electroneutrality and electronegativity balance were employed together in a recent report.[1] Examples of other principles include radius ratio rules,[21] *etc*. that are implemented in packages designed to facilitate materials discovery.[22] and covalence in high-throughput materials design.[23]

This type of algorithmic matching to make chemical formulae has been proven effective in the realization of several novel materials that were subsequently verified in experiment. In fact, 11 novel semiconductors were reported in a single study using this method. [24]

Table 1: Number of possible element combinations with the principle of electroneutrality enforced.

| Components | 2 | 3 | 4 | 5 |
|---|---|---|---|---|
| **Electroneutral combinations for our 84 Oxidation States** | 178 | 1,913 | 27,073 | 268,795 |
| **Electroneutral combinations for the 157 Oxidation States** | 510 | 9,858 | 245,551 | 4,307,895 |
| **Electroneutral combinations for the 367 Oxidation States** | 3,445 | 179,659 | 10,016,212 | 431,839,459 |

Our results illustrate the enormous difference just a few extra oxidation states can make in predicting viable compounds, and highlight the inherent compromise between the desire to capture as much chemistry as possible and prioritizing only the most viable discoveries. We believe that the 84 representative oxidation states we report in this work is the optimal trade-off for the community and should allow for a significant acceleration of modern materials discovery.

Our results also go some way to validate the heuristic approach to pedagogy in the chemical sciences. There is an inherent bias in scientific publishing towards results that are outside expected behavior, these are considered more interesting and therefore worthy of greater impact or attention. However, in materials discovery and introductory education, focusing on the most reliable paradigms is crucial in anticipating the likely behavior of elements in the context of inorganic chemistry.

ASSOCIATED CONTENT

**Supporting Information**.

The following information is provided free of charge in a PDF format.

SI Table 1: the raw information used to plot Figure 1(a), i.e. the number of matches for each element and oxidation state combination from the ICSD.

SI Table 2: the 84 representative oxidation states recommended in this work in table form, used to plot Figure 1(b).

SI Resource: the python script used to calculate the numbers used in Table 1. Available in text form here or as an executable online at https://github.com/bmd-lab/electroneutral_match

AUTHOR INFORMATION

**Notes**

The authors declare no competing financial interests.

ACKNOWLEDGMENT

The authors thank Adam J. Jackson for useful discussion. L.A.B acknowledges support by a grant-in-aid for JSPS fellows (no. 26-04792), and Shanghai Municipal Science and Technology Commission Program (no.19010500500).

REFERENCES

(1) Davies, D. W.; Butler, K. T.; Jackson, A. J.; Morris, A.; Frost, J. M.; Skelton, J. M.; Walsh, A. Computational Screening of All Stoichiometric Inorganic Materials. *Chem* **2016**, *1* (4), 617–627. https://doi.org/10.1016/j.chempr.2016.09.010.
(2) Levinthal, Cyrus. How to Fold Graciously; Allerton House, Monticello, Illinois, 1969; pp 22–24.
(3) Gavezzotti, A. Are Crystal Structures Predictable? *Acc. Chem. Res.* **1994**, *27* (10), 309–314. https://doi.org/10.1021/ar00046a004.
(4) Karen, P. Oxidation State, A Long-Standing Issue! *Angew. Chem. Int. Ed.* **2015**, *54* (16), 4716–4726. https://doi.org/10.1002/anie.201407561.
(5) Catlow, C. R. A.; Stoneham, A. M. Ionicity in Solids. *J. Phys. C Solid State Phys.* **1983**, *16* (22), 4321–4338. https://doi.org/10.1088/0022-3719/16/22/010.
(6) Walsh, A.; Sokol, A. A.; Buckeridge, J.; Scanlon, D. O.; Catlow, C. R. A. Oxidation States and Ionicity. *Nat. Mater.* **2018**, *17* (11), 958–964. https://doi.org/10.1038/s41563-018-0165-7.
(7) Liu, C.; Nikolaev, S. A.; Ren, W.; Burton, L. A. Electrides: A Review. *J. Mater. Chem. C* **2020**. https://doi.org/10.1039/D0TC01165G.
(8) Wang, G.; Zhou, M.; Goettel, J. T.; Schrobilgen, G. J.; Su, J.; Li, J.; Schlöder, T.; Riedel, S. Identification of an Iridium-Containing Compound with a Formal Oxidation State of IX. *Nature* **2014**, *514* (7523), 475–477. https://doi.org/10.1038/nature13795.
(9) Yu, H. S.; Truhlar, D. G. Oxidation State 10 Exists. *Angew. Chem. Int. Ed.* **2016**, *55* (31), 9004–9006. https://doi.org/10.1002/anie.201604670.
(10) IUPAC - oxidation state (O04365) https://goldbook.iupac.org/terms/view/O04365 (accessed Jul 3, 2020). https://doi.org/10.1351/goldbook.O04365.
(11) Lin, Z.; Wang, L.; Zhang, J.; Mao, H.; Zhao, Y. Nanocrystalline Tungsten Carbide: As Incompressible as Diamond. *Appl. Phys. Lett.* **2009**, *95* (21), 211906. https://doi.org/10.1063/1.3268457.
(12) Kojima, A.; Teshima, K.; Shirai, Y.; Miyasaka, T. Organometal Halide Perovskites as Visible-Light Sensitizers for Photovoltaic Cells. *J. Am. Chem. Soc.* **2009**, *131* (17), 6050–6051. https://doi.org/10.1021/ja809598r.
(13) Lide, D. R. *CRC Handbook of Chemistry and Physics*; CRC Press: Boca Raton, 2005.
(14) Greenwood, N. N.; Earnshaw, A. *Chemistry of the Elements - 2nd Edition*; Elsevier.
(15) Hautier, G.; Fischer, C.; Ehrlacher, V.; Jain, A.; Ceder, G. Data Mined Ionic Substitutions for the Discovery of New Compounds. *Inorg. Chem.* **2011**, *50* (2), 656–663. https://doi.org/10.1021/ic102031h.
(16) Oganov, A. R.; Glass, C. W. Crystal Structure Prediction Using Ab Initio Evolutionary Techniques: Principles and Applications. *J. Chem. Phys.* **2006**, *124* (24), 244704. https://doi.org/10.1063/1.2210932.
(17) Pickard, C. J.; Needs, R. J. High-Pressure Phases of Silane. *Phys. Rev. Lett.* **2006**, *97* (4), 045504. https://doi.org/10.1103/PhysRevLett.97.045504.
(18) Podryabinkin, E. V.; Tikhonov, E. V.; Shapeev, A. V.; Oganov, A. R. Accelerating Crystal Structure Prediction by Machine-Learning Interatomic Potentials with Active Learning. *Phys. Rev. B* **2019**, *99* (6), 064114. https://doi.org/10.1103/PhysRevB.99.064114.

(19) Berger, R. F.; Neaton, J. B. Computational Design of Low-Band-Gap Double Perovskites. *Phys. Rev. B* **2012**, *86* (16), 165211. https://doi.org/10.1103/PhysRevB.86.165211.
(20) Serrate, D.; Teresa, J. M. D.; Ibarra, M. R. Double Perovskites with Ferromagnetism above Room Temperature. *J. Phys. Condens. Matter* **2006**, *19* (2), 023201. https://doi.org/10.1088/0953-8984/19/2/023201.
(21) Goldschmidt, V. M. Die Gesetze der Krystallochemie. *Naturwissenschaften* **1926**, *14* (21), 477–485. https://doi.org/10.1007/BF01507527.
(22) Davies, D. W.; Butler, K. T.; Jackson, A. J.; Skelton, J. M.; Morita, K.; Walsh, A. SMACT: Semiconducting Materials by Analogy and Chemical Theory. *J. Open Source Softw.* **2019**, *4* (38), 1361. https://doi.org/10.21105/joss.01361.
(23) Sun, W.; Bartel, C. J.; Arca, E.; Bauers, S. R.; Matthews, B.; Orvañanos, B.; Chen, B.-R.; Toney, M. F.; Schelhas, L. T.; Tumas, W.; Tate, J.; Zakutayev, A.; Lany, S.; Holder, A. M.; Ceder, G. A Map of the Inorganic Ternary Metal Nitrides. *Nat. Mater.* **2019**, *18* (7), 732–739. https://doi.org/10.1038/s41563-019-0396-2.
(24) Hinuma, Y.; Hatakeyama, T.; Kumagai, Y.; Burton, L. A.; Sato, H.; Muraba, Y.; Iimura, S.; Hiramatsu, H.; Tanaka, I.; Hosono, H.; Oba, F. Discovery of Earth-Abundant Nitride Semiconductors by Computational Screening and High-Pressure Synthesis. *Nat. Commun.* **2016**, *7* (1), 11962. https://doi.org/10.1038/ncomms11962.

# Supporting Information for Data-Mining Element Charges in Inorganic Materials

*Yu ding[a], Yu Kumagai[b], Fumiyasu Oba[b] and Lee A. Burton[a]\**

[a] International Centre for Quantum and Molecular Structures, Department of Physics, Shanghai University, Shanghai 200444, China.

[b] Laboratory for Materials and Structures, Institute of Innovative Research, Tokyo Institute of Technology, 4259 Nagatsuta, Midori-ku, Yokohama 226-8503, Japan.

AUTHOR INFORMATION

**Corresponding Author**

\* E-mail: leeburton@shu.edu.cn Fax: +86 21 66134208 ; Tel: +86 2166136129;

SI Table 1: Number of matches returned from ICSD repository shown in Figure 1(a) of main text.

| Element | -4 | -3 | -2 | -1 | +1 | +2 | +3 | +4 | +5 | +6 | +7 | +8 |
|---|---|---|---|---|---|---|---|---|---|---|---|---|
| H | 0 | 2 | 3 | 20556 | 67903 | 1 | 4 | 8 | 0 | 0 | 0 | 0 |
| He | 0 | 0 | 0 | 0 | 0 | 0 | 0 | 0 | 0 | 0 | 0 | 0 |
| Li | 1 | 0 | 0 | 0 | 12557 | 0 | 0 | 0 | 0 | 0 | 0 | 0 |
| Be | 0 | 0 | 0 | 0 | 0 | 1609 | 0 | 0 | 0 | 0 | 0 | 0 |
| B | 3 | 353 | 82 | 49 | 205 | 54 | 11625 | 0 | 0 | 0 | 0 | 0 |
| C | 5468 | 158 | 3767 | 371 | 927 | 16960 | 4625 | 19541 | 0 | 0 | 0 | 0 |
| N | 13 | 40140 | 1401 | 1813 | 753 | 215 | 1759 | 49 | 3576 | 1 | 0 | 0 |
| O | 16 | 2 | 694430 | 3580 | 0 | 0 | 5 | 0 | 0 | 3 | 1 | 0 |
| F | 4 | 2 | 4 | 46069 | 0 | 2 | 0 | 0 | 0 | 0 | 0 | 0 |
| Ne | 0 | 0 | 0 | 0 | 0 | 0 | 0 | 0 | 0 | 0 | 0 | 0 |
| Na | 0 | 0 | 0 | 2 | 33104 | 0 | 0 | 0 | 0 | 0 | 0 | 0 |
| Mg | 0 | 0 | 0 | 1 | 0 | 18197 | 0 | 0 | 0 | 0 | 0 | 0 |
| Al | 2 | 2 | 0 | 0 | 1 | 9 | 28079 | 0 | 0 | 0 | 0 | 0 |
| Si | 347 | 33 | 102 | 79 | 9 | 96 | 69 | 36988 | 0 | 0 | 0 | 0 |
| P | 16 | 2040 | 259 | 390 | 146 | 101 | 1379 | 861 | 24432 | 2 | 0 | 0 |
| S | 1 | 1 | 45357 | 2061 | 95 | 494 | 151 | 2352 | 177 | 8274 | 0 | 3 |
| Cl | 8 | 0 | 7 | 29346 | 4 | 4 | 60 | 12 | 150 | 0 | 644 | 0 |
| Ar | 0 | 0 | 0 | 0 | 0 | 0 | 0 | 0 | 0 | 0 | 0 | 0 |
| K | 0 | 0 | 0 | 0 | 30529 | 0 | 0 | 0 | 0 | 0 | 0 | 0 |
| Ca | 0 | 0 | 0 | 6 | 3 | 21620 | 0 | 0 | 0 | 0 | 0 | 0 |
| Sc | 0 | 0 | 0 | 0 | 36 | 38 | 1795 | 0 | 0 | 0 | 0 | 0 |
| Ti | 0 | 0 | 0 | 0 | 5 | 220 | 789 | 12191 | 0 | 0 | 0 | 0 |
| V | 0 | 0 | 0 | 0 | 16 | 194 | 1509 | 2613 | 5405 | 1 | 0 | 0 |
| Cr | 0 | 0 | 0 | 0 | 11 | 597 | 4364 | 213 | 174 | 965 | 0 | 0 |
| Mn | 0 | 1 | 0 | 24 | 95 | 9504 | 3188 | 1399 | 53 | 15 | 89 | 0 |
| Fe | 0 | 0 | 1 | 13 | 119 | 10580 | 14439 | 222 | 6 | 8 | 0 | 0 |
| Co | 0 | 0 | 1 | 45 | 93 | 5655 | 2215 | 102 | 0 | 0 | 0 | 0 |
| Ni | 0 | 0 | 3 | 0 | 102 | 5319 | 652 | 84 | 0 | 0 | 0 | 0 |
| Cu | 0 | 0 | 1 | 0 | 6773 | 8924 | 375 | 2 | 0 | 0 | 0 | 0 |
| Zn | 0 | 0 | 3 | 5 | 0 | 12938 | 13 | 3 | 0 | 0 | 0 | 0 |
| Ga | 2 | 12 | 2 | 2 | 47 | 124 | 6471 | 11 | 1 | 0 | 0 | 0 |
| Ge | 201 | 37 | 68 | 69 | 2 | 286 | 125 | 5104 | 0 | 0 | 0 | 0 |
| As | 5 | 2567 | 286 | 285 | 67 | 330 | 2435 | 26 | 4349 | 0 | 0 | 0 |
| Se | 0 | 0 | 17660 | 1244 | 64 | 138 | 18 | 2388 | 6 | 1310 | 0 | 2 |
| Br | 0 | 1 | 0 | 10322 | 23 | 0 | 27 | 0 | 178 | 0 | 32 | 0 |
| Kr | 0 | 0 | 0 | 0 | 0 | 35 | 0 | 0 | 0 | 0 | 0 | 0 |
| Rb | 0 | 0 | 0 | 2 | 8567 | 0 | 0 | 0 | 0 | 0 | 0 | 0 |
| Sr | 0 | 0 | 0 | 0 | 0 | 14460 | 0 | 0 | 0 | 0 | 0 | 0 |
| Y | 0 | 0 | 0 | 0 | 14 | 51 | 5468 | 16 | 0 | 0 | 0 | 0 |
| Zr | 0 | 0 | 0 | 0 | 28 | 97 | 110 | 4398 | 0 | 0 | 0 | 0 |
| Nb | 0 | 0 | 0 | 0 | 18 | 92 | 315 | 576 | 6940 | 3 | 0 | 0 |
| Mo | 0 | 0 | 0 | 0 | 18 | 409 | 455 | 549 | 1947 | 11523 | 6 | 0 |
| Tc | 0 | 0 | 0 | 0 | 13 | 15 | 80 | 56 | 7 | 1 | 73 | 0 |
| Ru | 0 | 0 | 0 | 0 | 14 | 310 | 258 | 774 | 533 | 93 | 1 | 6 |
| Rh | 0 | 0 | 0 | 6 | 27 | 81 | 549 | 104 | 12 | 0 | 0 | 0 |
| Pd | 0 | 0 | 0 | 0 | 88 | 1452 | 31 | 95 | 0 | 0 | 0 | 0 |
| Ag | 0 | 0 | 0 | 0 | 8212 | 151 | 36 | 0 | 0 | 0 | 0 | 0 |
| Cd | 0 | 0 | 0 | 0 | 12 | 6591 | 2 | 0 | 0 | 0 | 0 | 0 |
| **In** | 0 | 3 | 0 | 17 | 475 | 201 | 4967 | 0 | 0 | 0 | 0 | 0 |
| **Sn** | 50 | 28 | 51 | 61 | 0 | 1746 | 69 | 3307 | 0 | 0 | 0 | 0 |
| **Sb** | 0 | 1224 | 245 | 154 | 1 | 18 | 4906 | 82 | 2660 | 1 | 0 | 0 |
| **Te** | 0 | 1 | 5905 | 912 | 66 | 112 | 0 | 2414 | 13 | 1292 | 0 | 0 |
| **I** | 3 | 3 | 0 | 11965 | 73 | 0 | 46 | 6 | 1365 | 4 | 243 | 0 |
| **Xe** | 0 | 0 | 0 | 0 | 0 | 226 | 4 | 13 | 0 | 142 | 0 | 6 |
| **Cs** | 0 | 0 | 0 | 0 | 12766 | 0 | 0 | 0 | 0 | 0 | 0 | 0 |
| **Ba** | 0 | 0 | 0 | 0 | 0 | 18494 | 5 | 0 | 0 | 0 | 0 | 0 |
| **La** | 0 | 0 | 0 | 0 | 7 | 95 | 11683 | 8 | 0 | 0 | 0 | 0 |
| **Ce** | 0 | 0 | 0 | 4 | 0 | 87 | 3313 | 1043 | 0 | 0 | 0 | 0 |
| **Pr** | 0 | 0 | 0 | 0 | 0 | 51 | 3213 | 114 | 0 | 0 | 0 | 0 |
| **Nd** | 0 | 0 | 0 | 0 | 0 | 97 | 4899 | 11 | 1 | 0 | 0 | 0 |
| **Pm** | 0 | 0 | 0 | 0 | 0 | 0 | 13 | 0 | 0 | 0 | 0 | 0 |
| **Sm** | 0 | 0 | 0 | 0 | 1 | 165 | 2464 | 5 | 0 | 0 | 0 | 0 |
| **Eu** | 0 | 0 | 0 | 0 | 2 | 1230 | 1712 | 7 | 0 | 0 | 0 | 0 |
| **Gd** | 0 | 0 | 0 | 0 | 3 | 49 | 2866 | 6 | 0 | 0 | 0 | 0 |
| **Tb** | 0 | 0 | 0 | 0 | 11 | 34 | 1309 | 104 | 0 | 0 | 0 | 0 |
| **Dy** | 0 | 0 | 0 | 0 | 0 | 59 | 1712 | 3 | 0 | 0 | 0 | 0 |
| **Ho** | 0 | 0 | 0 | 0 | 0 | 18 | 1702 | 2 | 0 | 0 | 0 | 0 |
| **Er** | 0 | 0 | 0 | 0 | 0 | 28 | 1971 | 3 | 0 | 0 | 0 | 0 |
| **Tm** | 0 | 0 | 0 | 0 | 0 | 37 | 943 | 2 | 0 | 0 | 0 | 0 |
| **Yb** | 0 | 0 | 0 | 0 | 0 | 281 | 1734 | 4 | 0 | 0 | 0 | 0 |
| **Lu** | 0 | 0 | 0 | 0 | 0 | 5 | 1162 | 2 | 0 | 0 | 0 | 0 |
| **Hf** | 0 | 0 | 0 | 0 | 2 | 26 | 48 | 820 | 0 | 0 | 0 | 0 |
| **Ta** | 0 | 0 | 0 | 0 | 40 | 41 | 121 | 289 | 3112 | 3 | 0 | 0 |
| **W** | 0 | 0 | 0 | 0 | 1 | 116 | 45 | 188 | 203 | 34388 | 0 | 0 |
| **Re** | 0 | 0 | 0 | 0 | 75 | 30 | 1143 | 357 | 214 | 200 | 613 | 0 |
| **Os** | 0 | 0 | 0 | 0 | 30 | 78 | 59 | 121 | 93 | 80 | 21 | 61 |
| **Ir** | 0 | 0 | 1 | 3 | 15 | 57 | 256 | 423 | 137 | 18 | 0 | 0 |
| **Pt** | 0 | 0 | 6 | 0 | 4 | 644 | 99 | 650 | 11 | 3 | 0 | 0 |
| **Au** | 0 | 0 | 0 | 54 | 689 | 52 | 633 | 0 | 22 | 0 | 0 | 0 |
| **Hg** | 0 | 0 | 0 | 0 | 492 | 2857 | 0 | 0 | 0 | 0 | 0 | 0 |
| **Tl** | 0 | 0 | 3 | 68 | 3407 | 14 | 1093 | 0 | 0 | 0 | 0 | 0 |
| **Pb** | 8 | 0 | 0 | 11 | 4 | 9485 | 5 | 412 | 0 | 0 | 0 | 0 |
| **Bi** | 0 | 151 | 42 | 17 | 157 | 130 | 10147 | 66 | 270 | 0 | 0 | 0 |
| **Po** | 0 | 0 | 2 | 0 | 0 | 1 | 0 | 6 | 0 | 0 | 0 | 0 |
| **At** | 0 | 0 | 0 | 0 | 0 | 0 | 0 | 0 | 0 | 0 | 0 | 0 |
| **Rn** | 0 | 0 | 0 | 0 | 0 | 0 | 0 | 0 | 0 | 0 | 0 | 0 |
| **Fr** | 0 | 0 | 0 | 0 | 0 | 0 | 0 | 0 | 0 | 0 | 0 | 0 |
| **Ra** | 0 | 0 | 0 | 0 | 0 | 3 | 0 | 0 | 0 | 0 | 0 | 0 |
| **Ac** | 0 | 0 | 0 | 0 | 0 | 0 | 7 | 0 | 0 | 0 | 0 | 0 |
| **Th** | 0 | 0 | 0 | 0 | 0 | 29 | 66 | 959 | 5 | 4 | 0 | 0 |
| **Pa** | 0 | 1 | 0 | 0 | 0 | 1 | 5 | 21 | 28 | 0 | 0 | 0 |
| **U** | 0 | 0 | 0 | 2 | 1 | 66 | 325 | 1002 | 160 | 4382 | 0 | 0 |
| **Np** | 0 | 0 | 0 | 1 | 0 | 6 | 40 | 114 | 176 | 59 | 21 | 0 |
| **Pu** | 0 | 0 | 0 | 0 | 0 | 13 | 89 | 92 | 3 | 30 | 7 | 0 |
| **Am** | 0 | 0 | 0 | 0 | 0 | 6 | 47 | 11 | 2 | 1 | 0 | 0 |
| **Cm** | 0 | 0 | 0 | 0 | 0 | 9 | 61 | 6 | 0 | 0 | 0 | 0 |
| **Bk** | 0 | 0 | 0 | 0 | 0 | 1 | 13 | 5 | 0 | 0 | 0 | 0 |
| **Cf** | 0 | 0 | 0 | 0 | 0 | 1 | 20 | 1 | 0 | 0 | 0 | 0 |
| **Es** | 0 | 0 | 0 | 0 | 0 | 0 | 2 | 0 | 0 | 0 | 0 | 0 |
| **Fm** | 0 | 0 | 0 | 0 | 0 | 0 | 0 | 0 | 0 | 0 | 0 | 0 |
| **Md** | 0 | 0 | 0 | 0 | 0 | 0 | 0 | 0 | 0 | 0 | 0 | 0 |
| **No** | 0 | 0 | 0 | 0 | 0 | 0 | 0 | 0 | 0 | 0 | 0 | 0 |
| **Lr** | 0 | 0 | 0 | 0 | 0 | 0 | 0 | 0 | 0 | 0 | 0 | 0 |
| **Rf** | 0 | 0 | 0 | 0 | 0 | 0 | 0 | 0 | 0 | 0 | 0 | 0 |
| **Db** | 0 | 0 | 0 | 0 | 0 | 0 | 0 | 0 | 0 | 0 | 0 | 0 |
| **Sg** | 0 | 0 | 0 | 0 | 0 | 0 | 0 | 0 | 0 | 0 | 0 | 0 |
| **Bh** | 0 | 0 | 0 | 0 | 0 | 0 | 0 | 0 | 0 | 0 | 0 | 0 |
| **Hs** | 0 | 0 | 0 | 0 | 0 | 0 | 0 | 0 | 0 | 0 | 0 | 0 |

SI Table 2: Most suitable representative oxidation states (OS) presented in Figure 1(b) of main text.

| Element | Chosen OS | Element | Chosen OS | Element | Chosen Os | Element | Chosen OS |
|---|---|---|---|---|---|---|---|
| **H** | +1 | **Cr** | +3 | **Pd** | +2 | **Ho** | +3 |
| **Li** | +1 | **Mn** | +2 | **Ag** | +1 | **Er** | +3 |
| **Be** | +2 | **Fe** | +2,+3 | **Cd** | +2 | **Tm** | +3 |
| **B** | +3 | **Co** | +2 | **In** | +3 | **Yb** | +3 |
| **C** | -4 | **Ni** | +2 | **Sn** | +2,+4 | **Lu** | +3 |
| **N** | -3 | **Cu** | +1,+2 | **Sb** | +3 | **Hf** | +4 |
| **O** | -2 | **Zn** | +2 | **Te** | -2 | **Ta** | +5 |
| **F** | -1 | **Ga** | +3 | **I** | -1 | **W** | +6 |
| **Na** | +1 | **Ge** | +4 | **Cs** | +1 | **Re** | +3 |
| **Mg** | +2 | **As** | -3,+5 | **Ba** | +2 | **Os** | +4 |
| **Al** | +3 | **Se** | -2 | **La** | +3 | **Ir** | +4 |
| **Si** | +4 | **Br** | -1 | **Ce** | +3 | **Pt** | +2,+4 |
| **P** | +5 | **Rb** | +1 | **Pr** | +3 | **Au** | +1,+3 |
| **S** | -2 | **Sr** | +2 | **Nd** | +3 | **Hg** | +2 |
| **Cl** | -1 | **Y** | +3 | **Pm** | +3 | **Tl** | +1 |
| **K** | +1 | **Zr** | +4 | **Sm** | +3 | **Pb** | +2 |
| **Ca** | +2 | **Nb** | +5 | **Eu** | +2,+3 | **Bi** | +3 |
| **Sc** | +3 | **Mo** | +6 | **Gd** | +3 | | |
| **Ti** | +4 | **Ru** | +4 | **Tb** | +3 | | |
| **V** | +5 | **Rh** | +3 | **Dy** | +3 | | |

SI Resource 1: Python code used to calculate electroneutral combinations shown in Table 1 of main text.

```python
#! /usr/bin/env python3
from itertools import combinations_with_replacement
from typing import Sequence

n_ions = 5
data_file = "file_of_oxidation_states"

def all_electroneutral_combinations(charges: Sequence[int], n_ions: int) -> list:
    """Given a list of charges, return all the neutral combinations.
    Combinations include repetition (i.e. for charges=[-1, 2], n_ions=3
    [-1, -2, 2] would be found as a neutral combination)

    Args:
        charges: Sequence of charges to consider
        n_ions: number of charges to combine in each considered combination

    Returns:
        List of neutral combinations from input charges
    """

    return [x for x in combinations_with_replacement(charges, n_ions)
            if sum(x) == 0]

with open(data_file, 'rt') as fd:
    charges_from_file = [int(line.rstrip('\n')) for line in fd]

print("Neutral combinations: {}".format(
    len(all_electroneutral_combinations(charges_from_file, n_ions))))
```